\documentclass{article}

\usepackage{arxiv}

\usepackage[utf8]{inputenc} % allow utf-8 input
\usepackage[T1]{fontenc}    % use 8-bit T1 fonts
\usepackage{hyperref}       % hyperlinks
\usepackage{url}            % simple URL typesetting
\usepackage{booktabs}       % professional-quality tables
\usepackage{amsfonts}       % blackboard math symbols
\usepackage{nicefrac}       % compact symbols for 1/2, etc.
\usepackage{microtype}      % microtypography
\usepackage{lipsum}		% Can be removed after putting your text content
\usepackage{graphicx}
\usepackage[square,numbers]{natbib}
\usepackage{doi}
\usepackage{comment} %write comments
\usepackage{multirow}   %Use merge cell in tables
\usepackage{subcaption}
\usepackage{amsmath}  		% Better maths support

\title{Quantum Key Storage for Efficient Key Management}

\author{ \href{https://orcid.org/0000-0002-7981-7739}{\includegraphics[scale=0.06]{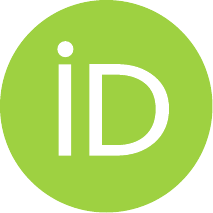}\hspace{1mm}Emir Dervisevic} \\
	Department of Telecommunications \\ Faculty of Electrical Engineering \\ 
        University of Sarajevo \\
	Sarajevo, Bosnia and Herzegovina \\
	\texttt{emir.dervisevic@etf.unsa.ba} \\
	\And
	{Amina Tankovic} \\
	Department of Telecommunications \\ Faculty of Electrical Engineering \\ 
        University of Sarajevo \\
	Sarajevo, Bosnia and Herzegovina \\
        \And
        {Enio Kaljic} \\
        Department of Telecommunications \\ Faculty of Electrical Engineering \\ 
        University of Sarajevo \\
	Sarajevo, Bosnia and Herzegovina \\
        \And 
        {Miroslav Voznak} \\
        VSB – Technical University of Ostrava \\
        Ostrava, Czechia \\
        \And
        {Miralem Mehic} \\
        Department of Telecommunications \\ Faculty of Electrical Engineering \\ 
        University of Sarajevo \\
	Sarajevo, Bosnia and Herzegovina \\
}

\renewcommand{\shorttitle}{Quantum Key Storage for Efficient Key Management}

\hypersetup{
pdftitle={Quantum Key Storage for Efficient Key Management},
pdfsubject={quantum key distribution, key management},
pdfauthor={Emir Dervisevic, Amina Tankovic, Enio Kaljic, Miroslav Voznak, Miralem Mehic},
pdfkeywords={Quantum Key Distribution, Storage, Key Management, Simulation},
}

\begin{document}
\maketitle

\begin{abstract}
    In the ongoing discourse surrounding integrating QKD networks as a service for critical infrastructures, key storage design often receives insufficient attention. Nonetheless, it bears crucial significance as it profoundly impacts the efficiency of QKD network services, thereby shaping its suitability for diverse applications. In this article, we analyze the effectiveness of key storage designs developed through practical testbeds and propose a novel key storage design to increase the effectiveness of key creation and supply. All key storage designs underwent analysis using network simulation tools, and the findings demonstrate that the novel key storage design surpasses existing approaches in terms of performance.
\end{abstract}

% keywords can be removed
\keywords{Quantum Key Distribution \and Key Management \and Key Storage \and Simulations}

\section{Introduction}
\label{sec:introduction}
Security has consistently remained paramount in light of the continual advancement of telecommunication services. Security mechanisms, widely employed and safeguarding communication for decades, may soon face insecurity~\cite{furrer2020roger, Mitchell2020}. The ongoing progress of both classical and quantum computing continually challenges their computational security. For these reasons, the development and implementation of quantum secure mechanisms are crucial~\cite{mehic2023quantum}. One of the most intriguing mechanisms, owing to its promising Information-Theoretic security, is Quantum Key Distribution (QKD)~\cite{bennett1984quantum}. It serves as a secret-key agreement primitive, facilitating the establishment of a secret key between two distant users via insecure channels. However, it has its challenges and costs. The limited distance and key generation rate are the primary challenges that diminish its practical applicability.\footnote{~In practical QKD networks, the secret key rate typically reaches only 1 $\sim$ 2 Mbps for a fiber distance of 50 km~\cite{dixon2010continuous, dynes2016ultra}.} The distance limitation is addressed by deploying trusted-repeater nodes~\cite{elliott2002building}. End nodes beyond a single QKD link use trusted-repeater nodes to establish keys in a process known as key relaying. However, this solution comes at the expense of expanding the trust circle to include intermediate nodes rather than solely end nodes. Because it can be challenging to ensure the trustworthiness of repeaters, methods have been developed to maintain security even when a certain number of nodes are compromised~\cite{rass2006achieving, rass2008secure, beals2008distributed, wen2009multiple}. This is accomplished by relaying parts of keys over non-intersecting routes in the QKD network. Another technique used to relax the trust assumption in the network is to perform key relaying so that the distributed key does not pass through intermediate nodes in clear-text~\cite{dong2019wide, vyas2024relaxing}.

\begin{figure}[!ht]
    \centering
    \includegraphics[width = 0.7 \textwidth]{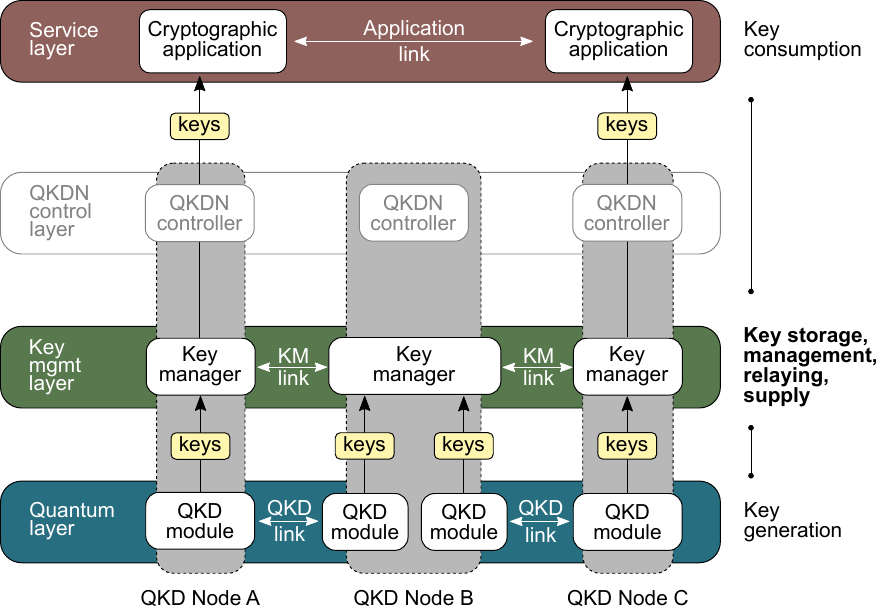}
    \caption{Layered network architecture for quantum key distribution. Quantum keys generated through the QKD process at the quantum layer are gathered and processed by the key management layer. Within this layer, the key manager is responsible for storing, managing, relaying, and supplying keys as needed. These keys are then utilized in the service layer to ensure secure data transmission between various applications.}
    \label{fig:architecture}
\end{figure}

QKD networks distinguish themselves from traditional networks by utilizing quantum and classical communication to generate, distribute, manage, and supply cryptographic keys. Consequently, QKD networks represent an add-on technology to telecommunications networks, providing the capability to establish ITS cryptographic keys.
QKD necessitates specialized hardware and is significantly computational and communication complex. It takes several minutes to complete a QKD process and establish a cryptographic key.
As a result, it is necessary to generate keys in advance to provide them promptly upon request. Generated keys must be securely stored and managed. The key manager is responsible for secure key storage, key lifecycle management, key relaying, and key supply~\cite{itu3803}. It is one of the indispensable components of the QKD network architecture~\cite{itu3800}: the central layer (see Figure~\ref{fig:architecture}) that segregates the key generation process from the key consumption process. Cryptographic applications access keys through standardized interfaces such as ETSI GS QKD 014~\cite{etsi014} and ETSI GS QKD 004~\cite{etsi004} or via proprietary protocols like the Cisco SKIP protocol~\cite{CiscoSKIP}. Keys are subsequently employed within a standard security framework to establish quantum-secure flows of data~\cite{dervisevic2021}.

How the key manager stores and manages the keys is critical, as it directly impacts the effectiveness of the QKD network as a service. However, discussions regarding the realization of this functional requirement of key managers often fail to address the efficacy comprehensively, instead focusing primarily on achieving a functional system overall. As a result, this article offers an analysis of common approaches to key storage design. The analysis was performed using QKDNetSim, a large-scale QKD network simulator~\cite{dervisevic2024large}. Furthermore, a novel key storage design is proposed to improve the efficiency of key supply. The new design demonstrated significantly improved performance in efficiently fulfilling requests for keys of various sizes.

The article is structured as follows: Section~\ref{sec:related} outlines existing key storage designs that have emerged from practical testbeds. In Section~\ref{sec:key-storage}, a novel key storage design is proposed. Section~\ref{sec:methodology} details the simulation setup and the experiments devised to test the designs, along with the relevant metrics measured. The results are presented and analyzed in Section~\ref{sec:results}, followed by further discussion in Section~\ref{sec:discussion}. Finally, Section~\ref{sec:conclusion} concludes the study.

\section{Related Work}
\label{sec:related}
\begin{figure}[!ht]
    \centering
    \includegraphics[width = 0.6 \textwidth]{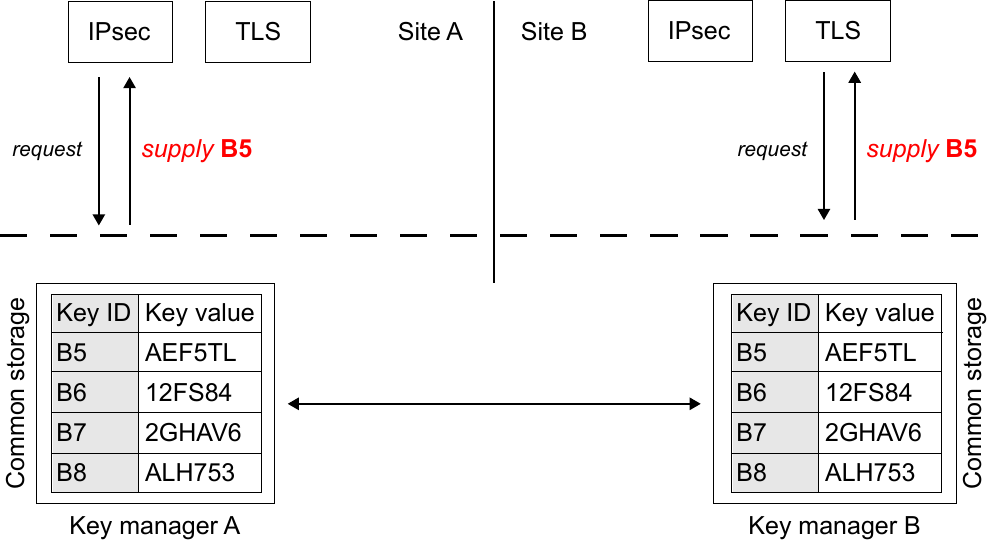}
    \caption{The key storage design with a single common storage. The two applications, IPsec and TLS, are running concurrently. In a short time frame, key managers serve the same key (B6) to different applications, resulting in key access collision.}
    \label{fig:collision}
\end{figure}

%talk about DARPA, SECOQC/TOSHIBA, NIST.
Over the decades--long period of QKD technology development, numerous practical implementations have been realized to showcase its applications~\cite{mehic2020quantum}. While aiming solely for a functional system, these practical implementations have inadvertently spurred the development of various approaches to key management and storage designs. 
The most straightforward storage design was introduced to develop the first QKD network -- the DARPA quantum network~\cite{elliott2002building, elliott2003quantum, elliott2007darpa}. The storage design comprises a single common storage where keys are stored in fixed sizes. To avoid key access collisions, which occur when two key managers simultaneously supply the same keys to different applications (see Figure~\ref{fig:collision}), a key reservation process must be performed at the key management layer (or at the service layer as done in the DARPA quantum network) before supply. This may result in slow operation and supply, which is unacceptable for applications running in critical infrastructures~\cite{tankovic2024performance}.

\begin{figure}[!ht]
    \centering
    \includegraphics[width = 0.7 \textwidth]{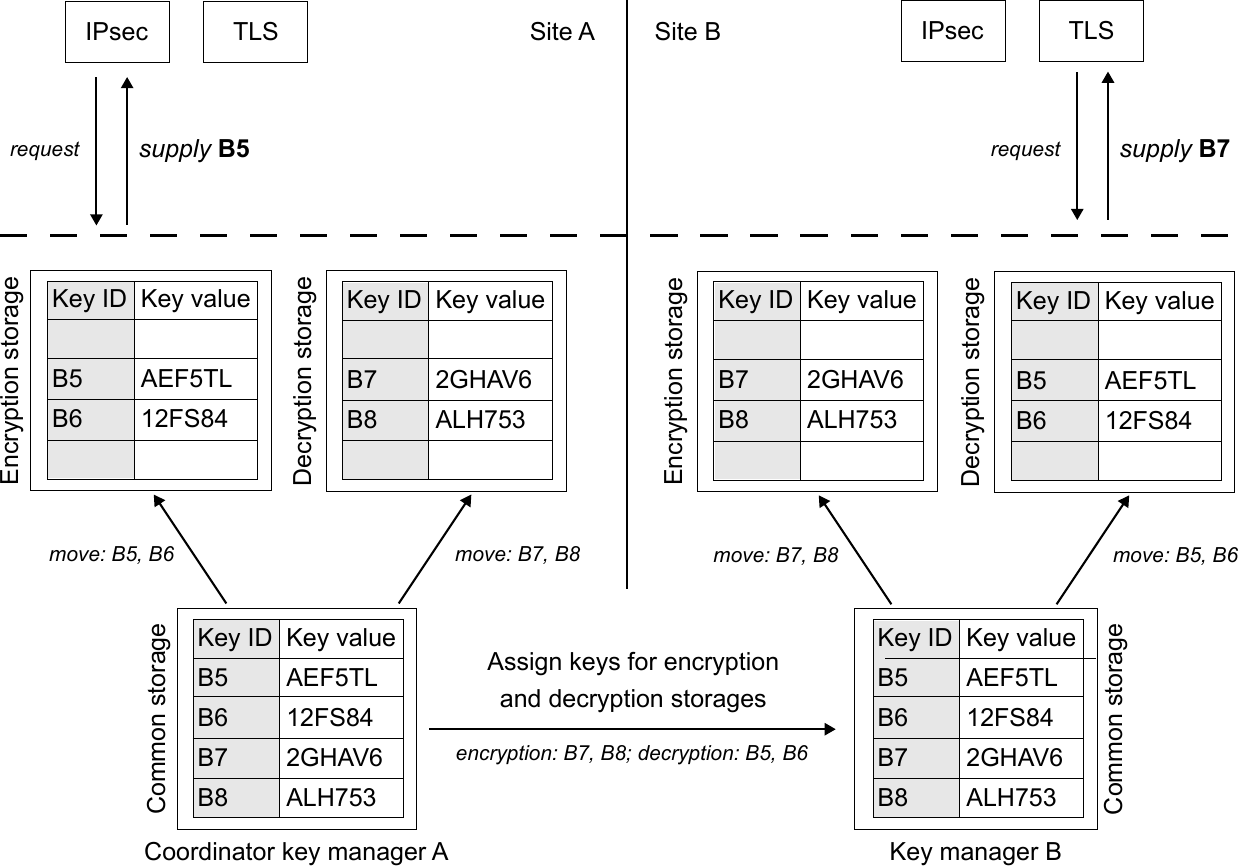}
    \caption{The key storage design that includes encryption and decryption key storages. Key managers assign keys for encryption and decryption purposes to achieve seamless key relaying and supply. The two applications, IPsec and TLS, are running concurrently. Key managers supply (use) keys from their respective encryption key storages. Multiple applications share access to encryption and decryption key storages.}
    \label{fig:enc-dec-design}
\end{figure}

\begin{figure}[!ht]
    \centering
    \includegraphics[width = 0.7 \textwidth]{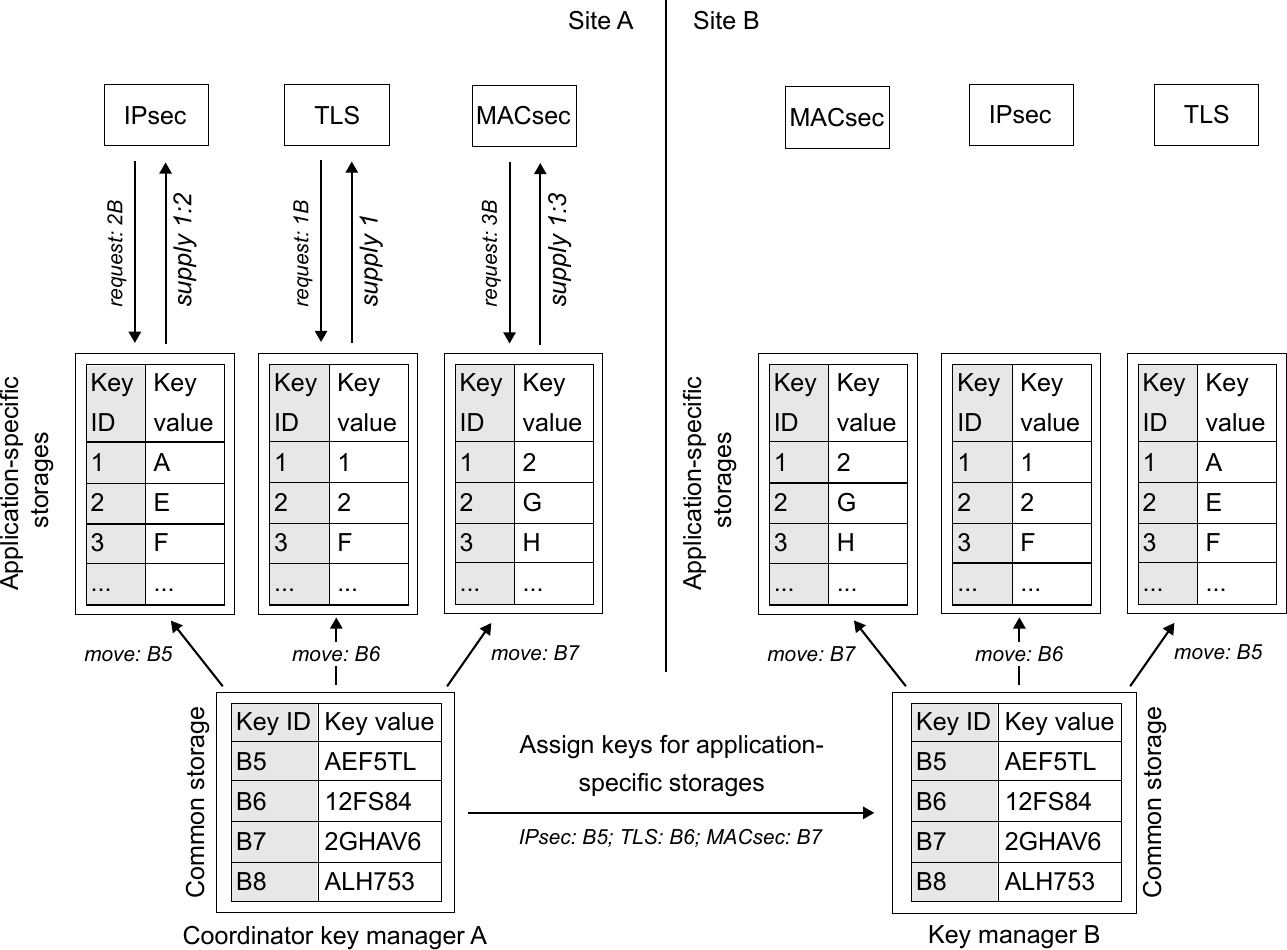}
    \caption{The key storage design that includes application-specific storages. Key managers allocate keys to designated storages. The three applications, IPsec, TLS and MACsec, are running concurrently. Key managers provide the requested number of bytes, ensuring the seamless supply of keys of varying sizes. The illustration depicts unidirectional traffic flow at the service layer, from site A to site B. This implies that applications perform encryption at site A using keys from their respective key storages. Application-specific storages at site B are accessed by corresponding receiving applications for decryption. For bidirectional traffic flow, applications establish a pair of storages for encryption and decryption.}
    \label{fig:app-specific-design}
\end{figure}

To prevent key access collisions and facilitate a seamless key relaying, keys are assigned specific purposes, such as encryption or decryption, at the key management layer~\cite{maurhart2010qkd, tanizawa2016secure, sasaki2011field}. This method ensures that key managers provide or use keys from their encryption key storage, as illustrated in Figure~\ref{fig:enc-dec-design}. When supplied, keys should undergo transformation processes such as merging, splitting, or combining both to align with the key size requirements of various cryptographic applications~\cite{dervisevic2024large}. Another key storage design, facilitating the straightforward provisioning of keys of different sizes, involves allocating dedicated key storage for each application~\cite{mink2008quantum}. Keys are reformatted to a size of one byte and stored in queues for each application as illustrated in Figure~\ref{fig:app-specific-design}. However, for applications with bidirectional traffic flow, separate storage must be established for each direction, leading to the creation of encryption and decryption key storage for every application. Both approaches, whether utilizing encryption and decryption or application-dedicated storage, necessitate one of the key managers to assume the master or coordinator role. The master oversees the timely and equitable distribution of available keys to these independent key storages.

Remarkably, despite the critical role of key storage designs in cryptography, there is a notable absence of comprehensive research that systematically analyzes the various approaches to this crucial aspect.

%%%%%%%%%%%%%%%%%%%%%%%%%%%%%%%%%%%%%%%%%%
\section{The Novel Key Storage Design for Efficient Key Management} 
\label{sec:key-storage}
\begin{figure}[!ht]
    \centering
    \includegraphics[width = 0.7 \textwidth]{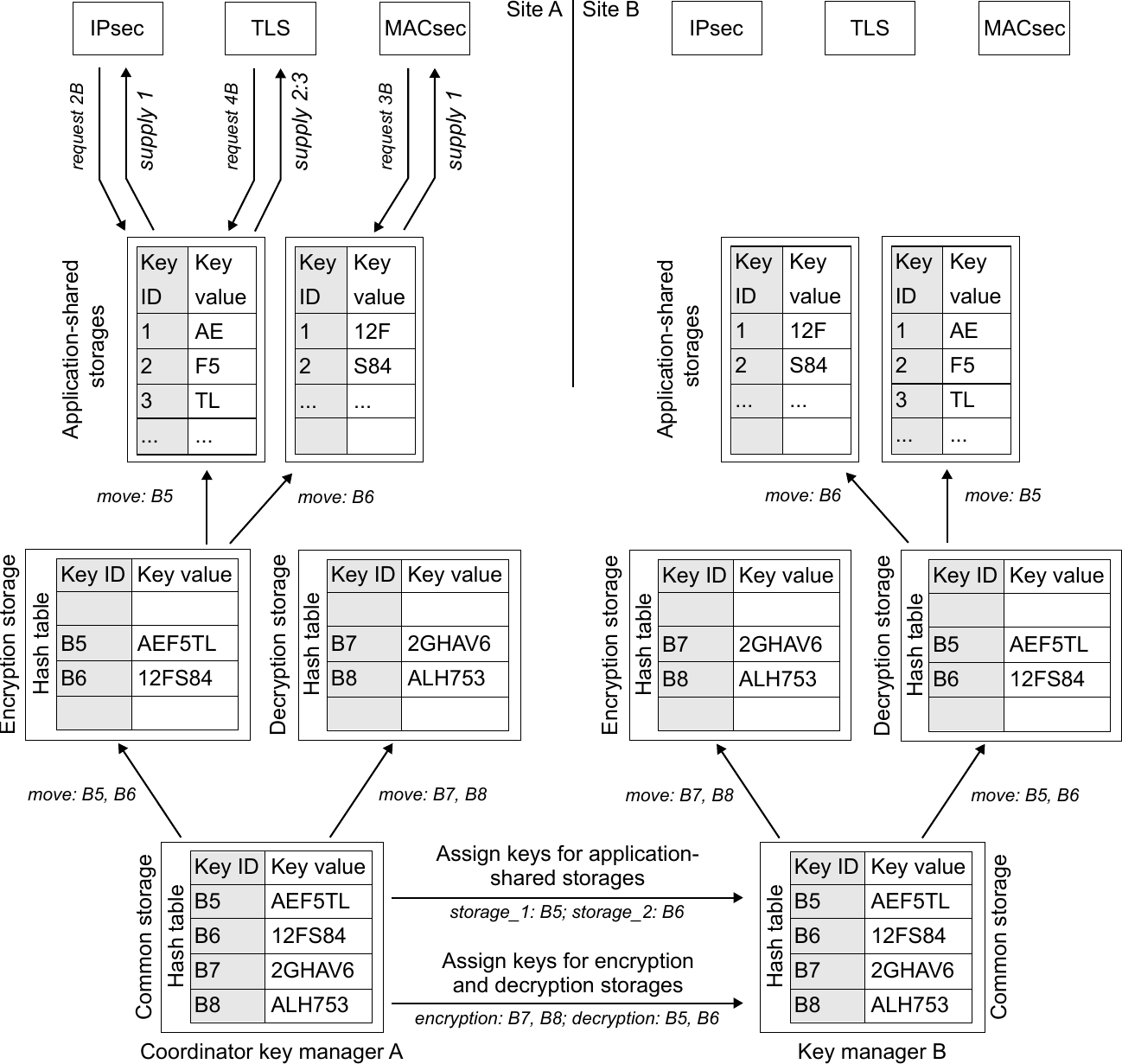}
    \caption{The novel key storage design that includes encryption and decryption key storages and modified application-specific storages. To avoid time-consuming and resource-intensive key transformation operations during supply, stores containing pre-formatted keys of specific sizes are created. The established storages are shared by multiple applications with similar key size requirements. For example, the IPsec and TLS applications have compatible key size requirements, so their requests can be fulfilled using a single shared storage. The supply key for TLS application can be effectively created pulling two keys from the storage shared with IPsec. The illustration depicts unidirectional traffic flow at the service layer, from site A to site B. This implies that applications perform encryption at site A using keys from their respective key storages. Application-shared storages at site B are accessed by corresponding receiving applications for decryption. For bidirectional traffic flow, applications establish a pair of storages for encryption and decryption.}
    \label{fig:novel-design}
\end{figure}

As the goal of integrating QKD technology into an enterprise environment progresses, it is critical to recognize that one key manager will serve thousands of clients with  diverse needs. To ensure that the technology is suitable for critical infrastructures, keys of varying sizes must be efficiently supplied on demand. The effectiveness of the key manager is determined by the design of the key storage structure. As a result, it is critical to find a key storage design that enables effective key supply. This section addresses this issue by introducing a design for effective key management that combines the design of encryption and decryption key storages with a modified application-dedicated key storage design, where multiple applications share storages. Figure~\ref{fig:novel-design} depicts the proposed key storage design. The basic idea is to create a key storage for specific key sizes required by applications. In this manner, the process of transforming the keys is avoided or reduced to a simple one that requires merging fewer keys. As a result, supply keys of the desired sizes can be efficiently created and supplied on demand. The reason for retaining encryption and decryption key storages in this novel design will be discussed in Section~\ref{sec:discussion}.

The following subsections elaborate on the requirements for a fast and reliable design, considering the choice of data structure, common storage, encryption and decryption storages, and application-shared storage.

%\textbf{Ovdje nedostaje jedna rečenica/paragraf koji objašnjava potrebu za uvođenjem narednih podsekcija. Bez te rečenice nejasno je zašto niste samo nastavili pisati sve u jednoj sekciji, zašto "sjeckate" tekst u nekoliko podsekcija. Npr. Following subsections provide details about requirements for fast and reliable key organization consdering common storage, encryption and decryption storage and application-shared storage.}

%\textbf{NAkon čitanja cijelog rada, smatram da bi bilo dobro na početu ovog rada, tj. ovdje grafički prikazati svaku od opisanih design pristupa. Rad zaista ima dobrog potencijala, razmišljam o vrhunskim časopisima, ali nedostaju detalji da se "rad ispolira".}

\begin{comment}
Potencijalni časopisi:
\begin{itemize}
\item 
IEEE Transactions on Knowledge and Data Engineering
\url{https://ieeexplore.ieee.org/xpl/RecentIssue.jsp?punumber=69}

\item IEEE Transactions on Network and Service Management
\url{https://www.comsoc.org/publications/journals/ieee-tnsm}
\end{itemize}
\end{comment}

\subsection{Selection of a suitable data structure}
\label{subsec:data-structure}
The method of organizing keys within key storage must be considered. Keys can be arranged in a sequential or unordered manner. These types are commonly referred to as linear and nonlinear data structures. Assigning keys to linear data structures makes the reformat function more convenient. As a result, keys can be organized in bytes or multiple-byte blocks. Examples are application-specific (see Figure~\ref{fig:app-specific-design}) and application-shared key storages (see Figure~\ref{fig:novel-design}). This design can provide a straightforward key supply compared to nonlinear data structures. 
However, the keys are ordered one after the other in linear data structures. If this operation is preceded by losing one or more bits, the key storage is desynchronized. Only the service layer can identify keys that are out of sync. The warning must be sent to the QKD network, purging and rebuilding the key storage. This is due to the difficulty of locating a resynchronization point in a random bit stream.
In contrast, nonlinear data structures store keys independently of one another. If one or more bits are lost in one key, the synchronization and correctness of the remaining keys are unaffected. Examples that use nonlinear data structures are common, encryption and decryption storages. The rationale behind the choice of data structure will become more evident and justified as we delve into the discussion of the design's key storages in the following subsections.

\subsection{The common storage}
\label{subsec:common-store}
Common storage is essential in any key storage design. While it's theoretically possible to omit this storage by assigning keys a purpose upon generation for encryption or decryption, such an approach is likely inefficient. Consequently, quantum keys received from QKD devices are stored in a common storage dedicated to the respective QKD link. Typically, this common storage should utilize non-volatile memory, often implemented using a database. Maintaining key storage, even in the event of a failure at the key manager device, is crucial. Non-volatile memory, such as a database, provides resilience against such failures by retaining stored keys even during system outages or crashes. The common storage serves as a reservoir for accumulating keys during periods of low demand. This accumulation is vital for meeting increased demand later on, ensuring sufficient keys are available to exceed the key generation rate of a link for a specific duration. This capability helps maintain the system's smooth operation even during fluctuating traffic patterns or sudden spikes in demand. As a result, maintaining such a key reservoir is essential for uninterrupted operation.

In our simulation model (described in Section~\ref{sec:methodology}), we use unordered\_map, a non-linear data structure, to realize the common storage, a Q-Buffer. The unordered\_map data structure is internally implemented as a hash table, allowing for high performance with each operation having an average constant-time complexity. This data structure is compatible with non-volatile memories, such as databases, making it appropriate for practical implementations of common storage. 

\subsection{The encryption and decryption storage}
\label{subsec:enc-dec-storage}
As sketched earlier, implementing encryption and decryption key storages aims to prevent conflicts in key access and ensure smooth key relaying. Keys from the common storage are selectively designated for encryption or decryption purposes based on requirements. These are volatile, working memories that allow quicker access but should be limited in the number of keys they store. They are provided with only sufficient keys from the common storage, which are then redistributed to application-shared storages (or supplied to applications after optional transformation, as described in Section~\ref{sec:discussion}). The master key manager ensures that an adequate number of keys are present in both encryption and decryption storages by assigning keys from the common storage in a timely and coordinated manner.

We use hash tables to store encryption and decryption keys, the unordered\_map in particular. The storages store keys in the same default key size as the common storage, so no key reformatting occurs as keys move from the common to encryption and decryption storages. Although it is conceivable to omit encryption and decryption key storages from the design, as we haven't assigned them a specific purpose at this stage, Section~\ref{sec:discussion} will explore why both encryption and decryption storages, as well as application-shared storages, are indispensable.\footnote{~Keys from the common storage could be assigned to designated application-shared storages, eliminating the need for the encryption and decryption storages implemented through hash tables.} %When a key of a specific size is requested, the key manager searches for encryption storage for candidate keys. Candidate keys are keys found in storage that are transformed to meet the request. While the requested key size is larger than the default key size, a random key from the storage is added to the candidate set. Otherwise, the storage is searched for the most appropriate key. This operation, typically a merge of multiple keys, is sometimes combined with the split operation. This is because the last candidate, the most appropriate key, is frequently spitted, with a precise remaining portion appended to the candidate's set while the other remains in storage. The peer key manager, which must search for candidate keys based on their identifiers, benefits from the performance advantages of hash tables. 

\subsection{The application-shared storage}
Multiple applications with the same key size requirements define and share a functional deque\footnote{~A deque, short for "double-ended queue," is a data structure that enables elements to be added to or removed from either the front or the back of the queue. While deque key storage operates in a first-in-first-out (FIFO) manner like a queue, it utilizes a deque instead due to the importance of accessing the last element when appending new key material.} key storage. The deques contain keys from the encryption storages, which are continually added to the back and supplied from the front. Keys are stored sequentially within the deque and reformatted to the requested size before storage. This pre-formatting allows for the efficient provision of keys of specific sizes from corresponding deques. Before supply, keys are assigned unique UUIDs. They are removed from the deque, and a message is sent over the KM link to synchronize supply key creation. Certain key sizes are served from smaller deques to minimize the number of deques required. The crucial factor is that the requested key size is evenly divisible by the deque key size. For instance, if the key manager has configured a deque to supply 256 bits, a request for a 512-bit key can be fulfilled by simply merging two 256-bit keys from the existing deque.
These features collectively enhance the efficiency of key supply while ensuring straightforward key management. 

\section{Methodology}
\label{sec:methodology}
We employ network simulation techniques to evaluate the proposed key storage structure's performance. Despite the emergence of expandable platforms and testbeds for network research, simulation remains crucial for scalability in terms of both size and speed of experimentation, repeatability, accelerated development, and education. This is especially true in the case of QKD networks, where deployments are highly costly and time-consuming. In this case, we use QKDNetSim\footnote{\url{https://www.qkdnetsim.info/}}~\cite{dervisevic2024large}, a large-scale QKD network simulator that focuses on public channel performance, key, and network management strategies. It encompasses major QKD node components, including QKD modules, key managers, network controllers, and cryptographic applications. The primary function of QKD modules is to create a cryptographic key by using quantum and authenticated classic communication. QKDNetSim mimics the latter -- classic communication via QKD Post-Processing Application. It is a simple application that sends data back and forth, generating network traffic similar to the QKD protocol. Established keys are delivered to key managers via HTTPS connections. Key managers format, synchronize, store, distribute, and provide keys to cryptographic applications upon request. Cryptographic applications are simple applications that use cryptographic keys to establish secure communication and transfer data regularly. They can use a combination of encryption and authentication algorithms, such as Advanced Encryption Standard (AES), OTP, Wegman-Carter, or SHA-512. Cryptographic applications and key managers can communicate using either ETSI-defined interfaces. Network controllers are simplified objects that receive a complete knowledge of network topology and calculate routes for key relaying based on the shortest distance. This information is provided to key managers so that key relaying can be performed hop--by--hop.

QKDNetSim divides keys by purpose to encryption and decryption storage and defines key storage for the key stream sessions. Keys received from the quantum layer are reformatted to a default key size and stored in a common storage space known as the Q--Buffer. The master key manager for this connection assigns keys from the Q-Buffer to encryption and decryption storages known as S-Buffers. QKDNetSim utilizes a non-linear data structure -- unordered\_map, for both common (i.e., Q--Buffer) and encryption/decryption (i.e., S--Buffers) storages. The unordered\_map data structure is internally implemented as a hash table, allowing for high performance with each operation having an average constant-time complexity. The S--Buffers contain keys that are merely transferred from the common storage while retaining their identifiers and sizes. 

To provide keys on request, keys can be transformed using merge and split operations or a combination. Transformation is used to meet the diverse requirements of cryptographic applications for keys of various sizes. This process is overly complex because it frequently necessitates merging and splitting operations. During a split operation, the remaining portion of the key in the storage is reserved and only used once the transformation is confirmed successful by the peer key manager. Because the same approach is used when relaying a key across a network, it is assumed that key managers within the network store keys in predefined, identical sizes. This assumption is made to keep key managers simple, as it would be necessary to transform keys (possibly at each hop) before relaying. This assumption, however, is not justified in large-scale QKD networks where multiple vendors are participating. The problem could be solved by formatting keys as they move from common storage to S--Buffers. %The STREAM S-Buffers have also been defined. They store keys in blocks of equal size to the chunk size agreed upon within the QoS metrics~\cite{mehic2022}.

The subsequent subsections describe the simulation setup for each experiment, providing a clear overview of the analyzed parameters and objectives. Each experiment is repeated multiple times (as specified for each experiment) with a deterministic seed of 100 and an increasing run number. The simulations were performed on a Dell Inspiron 5584 series laptop which is running on Intel Core i5-8265U processor and 8 GB of RAM. The operating system running is Ubuntu 22.04.4 LTS. Key storage designs (excluding a single common storage design) are compared in terms of the average CPU time required to create supply keys.  The CPU time for each GET\_KEY request is measured beginning with the request being read and ending with the creation and assignment of required supply keys and UUID values for supply. The simulation setup includes point-to-point connection, but the results are applicable to the global deployments of QKD networks due to such measurements. More details are provided in respective subsections.
%\textbf{Following subsections discuss simulation results of different storage designs. Each experiment was performed using RANDOM number of applications... (write here what is common to experiments, and are they related, have you used same SEED values for random numbers to make sure that values are unique between experiments... Some detail about simulation SETUP is needed here.}

\subsection{Single common storage design}
\label{subsec:met-single}

\subsubsection{Key access collisions versus residual key count}
\label{subsec:collision-count}

To illustrate the vulnerability of a single common store design, we conduct a simulation experiment to analyze the frequency of key access collisions in relation to the remaining quantity of keys within the storage. The key management system application has been modified to exclude S-Buffers, so keys are only served from the Q-Buffer. A simulation experiment consists of a single QKD link connecting two distant sites that host various concurrent applications. Table~\ref{tab:collision-count} displays the experiment-related configuration parameters. The default size for reformatting and storing keys received from the quantum layer is 64 bytes. Cryptographic applications send packets of 64 bytes encrypted with the One-Time Pad cipher. This means no transformation is required at the key management layer because the keys have already been stored in the requested size. 
Because applications generate unidirectional traffic flow, the term concurrent applications indicates an equal number of sender applications on each side. Furthermore, applications use the ETSI GS QKD 014 interface to communicate with the KMS and request a single key via each GET\_KEY query. The experiment is repeated 50 times. It is important to note that applications are started and stopped at the same times and have the same data rates in a particular run, which increases the likelihood of collisions. The variable settings simulate key access collisions on various residual keys in a common store. The number of key access collisions is expected to be heavily influenced by the amount of residual keys in the common store and the number and demands of active cryptographic applications. 
% \usepackage{multirow}
\renewcommand{\arraystretch}{1.3} % Default value: 1
\begin{table}[]
\centering
\caption{Simulation settings -- Key access collisions versus residual key count}
\resizebox{0.65 \textwidth}{!}{
\begin{tabular}{|l|l|c|}
\hline
\multirow{4}{*}{Quantum layer}        & Key rates {[}kbps{]}                                                   & 10, 50, 100, 500                     \\ \cline{2-3} 
                                      & Key sizes {[}bytes{]}                                                  & 1024, 2048, 4096, 8192, 16384, 32768 \\ \cline{2-3} 
                                      & Start {[}sec{]}                                                        & 0                                    \\ \cline{2-3} 
                                      & Stop {[}sec{]}                                                         & 70                                   \\ \hline
\multirow{2}{*}{Key management layer} & \begin{tabular}[c]{@{}l@{}}Default key size\\ {[}bytes{]}\end{tabular} & 64                                   \\ \cline{2-3} 
                                      & \begin{tabular}[c]{@{}l@{}}Storage capacity\\ {[}keys{]}\end{tabular}  & 100000                               \\ \hline
\multirow{7}{*}{Service layer}        & \begin{tabular}[c]{@{}l@{}}Number of\\ applications\end{tabular}       & 2, 4, 6, 8, 10, 12                   \\ \cline{2-3} 
                                      & Packet size {[}bytes{]}                                                & 64                                   \\ \cline{2-3} 
                                      & Data rates {[}kbps{]}                                                  & 1, 2, 3, 4, 5, 6, 7, 8, 9, 10        \\ \cline{2-3} 
                                      & Hold time {[}us{]}                                                     & 10                                   \\ \cline{2-3} 
                                      & Encryption                                                             & One-Time Pad                         \\ \cline{2-3} 
                                      & Start {[}sec{]}                                                        & 0, 1, 3, 5, 7, 9, 11, 13, 15         \\ \cline{2-3} 
                                      & Stop {[}sec{]}                                                         & 80, 85, 90, 95, 100                  \\ \hline
\end{tabular}
}
\label{tab:collision-count}
\end{table}
\renewcommand{\arraystretch}{1}

\subsubsection{Key access collisions versus application number}
\label{subsec:met-collision-application}
It is expected that for a given residual key count, the frequency of key access collisions will be influenced by the number of concurrent applications running. To explore this, a simulation experiment is devised, varying the number of cryptographic applications and assessing the performance of the quantum layer to accommodate their requirements. The key rate is defined as shown in Equation (\ref{eq:key-rate}), where $number$ is the number of concurrent applications running, $rate$ is their data rates, and $a$ is an increasing factor.

\begin{equation}
    \begin{aligned}
        Key \ rate = number \cdot rate \cdot (1 + a)
    \end{aligned}
    \label{eq:key-rate}
\end{equation}

This approach ensures that the residual key count remains low throughout the analysis. Table~\ref{tab:collision-num} displays the experiment-related configuration parameters. The experiment is repeated 1000 times. Similarly to the previous experiment, applications are started and stopped at the same times and with the same data rates during each run, increasing the likelihood of collisions.

% \usepackage{multirow}
\renewcommand{\arraystretch}{1.3} % Default value: 1
\begin{table}[]
\centering
\caption{Simulation settings -- Key access collisions versus application number}
\resizebox{0.6 \textwidth}{!}{
\begin{tabular}{|l|l|c|}
\hline
\multirow{4}{*}{Quantum layer}        & Key rates {[}kbps{]}         & Equation (1)               \\ \cline{2-3} 
                                      & Key sizes {[}bytes{]}        & 2048, 4096, 8192           \\ \cline{2-3} 
                                      & Start {[}sec{]}              & 0                          \\ \cline{2-3} 
                                      & Stop {[}sec{]}               & 100                        \\ \hline
\multirow{2}{*}{Key management layer} & Default key size {[}bytes{]} & 64                         \\ \cline{2-3} 
                                      & Storage capacity {[}keys{]}  & 100000                     \\ \hline
\multirow{7}{*}{Service layer}        & Number of applications       & 2, 4, 6, 8, 10, 20, 40, 50 \\ \cline{2-3} 
                                      & Packet size {[}bytes{]}      & 64                         \\ \cline{2-3} 
                                      & Data rates {[}kbps{]}        & 1, 2, 4                    \\ \cline{2-3} 
                                      & Hold time {[}ms{]}           & 100                        \\ \cline{2-3} 
                                      & Encryption                   & One-Time Pad               \\ \cline{2-3} 
                                      & Start {[}sec{]}              & 1                          \\ \cline{2-3} 
                                      & Stop {[}sec{]}               & 100                        \\ \hline
\end{tabular}
}
\label{tab:collision-num}
\end{table}
\renewcommand{\arraystretch}{1}

%In our second experiment, we compare the performance of the proposed key storage design to that implemented within QKDNetSim. To accomplish this, we implemented the key storage design shown in Figure X in QKDNetSim. The designs are compared in terms of time-complexity, RAM and CPU usage. 
\subsection{Key storage design that consists of encryption and decryption key storage}
\label{subsec:met-enc-dec}

\subsubsection{Encryption and decryption storage as a hash table}
\label{subsubsec:enc-dec-hash-table}
An experiment is designed to evaluate the performance of the key storage design, which consists of encryption and decryption storages implemented as hash tables. This design has already been implemented in QKDNetSim with Q and S--buffers. The performance is measured by the CPU time required to create a specified number of keys of a given size. Table~\ref{tab:storage-benchmark} displays the experiment-related configuration parameters. The elapsed CPU time is measured using $std::chrono::high\_resolution\_clock$ provided by the C++ standard library's $<chrono>$ header. It is designed to provide the highest possible resolution for timing measurements and is commonly used for measuring short durations accurately, such as performance profiling or benchmarking code. It is expected that the requested key size (or key number) and the S--Buffers default key size will impact the key creation time. For instance, the supply key could be made by just taking one key out of the encryption storage and giving it the UUID if the requested key size was the same as the default key size. In contrast, if the requested key size is smaller or larger than the default key size, multiple keys from the encryption storage must be merged or split. The experiment is repeated 1000 times. In contrast to previous experiments on key access collision, each application is assigned random (from a set) packet sizes, data rates, encryption algorithms (AES or OTP), number of keys fetched via a single GET\_KEY query, and start and stop times.

% \usepackage{multirow}
\renewcommand{\arraystretch}{1.3} % Default value: 1
\begin{table}[]
\centering
\caption{Simulation settings -- benchmarking supply key creation efficiency for different key storage designs}
\resizebox{0.65 \textwidth}{!}{
\begin{tabular}{|l|l|c|}
\hline
\multirow{4}{*}{Quantum layer}        & Key rates {[}kbps{]}                                                     & 10, 50, 100                           \\ \cline{2-3} 
                                      & Key sizes {[}bytes{]}                                                    & 1024, 2048, 4096, 8192, 16384, 32768  \\ \cline{2-3} 
                                      & Start {[}sec{]}                                                          & 0                                     \\ \cline{2-3} 
                                      & Stop {[}sec{]}                                                           & 100                                   \\ \hline
\multirow{2}{*}{Key management layer} & \begin{tabular}[c]{@{}l@{}}Default key size\\ {[}bytes{]}\end{tabular}   & 64                                    \\ \cline{2-3} 
                                      & \begin{tabular}[c]{@{}l@{}}Storage capacity\\ {[}keys{]}\end{tabular}    & 100000                                \\ \hline
\multirow{9}{*}{Service layer}        & \begin{tabular}[c]{@{}l@{}}Number of\\ applications\end{tabular}         & 2, 4, 6, 8, 10, 12, 20                \\ \cline{2-3} 
                                      & Packet size {[}bytes{]}                                                  & 100, 200, 300, 400, 500               \\ \cline{2-3} 
                                      & Data rates {[}kbps{]}                                                    & 1, 2, 3, 4, 5, 6, 7, 8, 9, 10         \\ \cline{2-3} 
                                      & Hold time {[}us{]}                                                       & 10                                    \\ \cline{2-3} 
                                      & Encryption                                                               & One-Time Pad, AES                     \\ \cline{2-3} 
                                      & Lifetime {[}bytes{]}                                                     & 1000, 5000, 10000, 50000              \\ \cline{2-3} 
                                      & \begin{tabular}[c]{@{}l@{}}Number of keys\\ in single query\end{tabular} & 1, 2, 3, 4, 5, 6                      \\ \cline{2-3} 
                                      & Start {[}sec{]}                                                          & 1, 2, 5, 7, 9, 11, 13, 15, 20, 40, 50 \\ \cline{2-3} 
                                      & Stop {[}sec{]}                                                           & 70, 75, 85, 90, 95, 100               \\ \hline
\end{tabular}
}
\label{tab:storage-benchmark}
\end{table}
\renewcommand{\arraystretch}{1}

\subsubsection{Encryption and decryption storage as a queue of bytes}
\label{subsubsec:enc-dec-queue}

Due to the appealing solution of dedicated-application storage presented in~\cite{mink2008quantum}, an experiment is designed to test its effectiveness. However, the storage design applies to encryption and decryption rather than application-specific storage. When requesting encryption keys, all applications can access a single storage queue. The performance benchmark, however, reveals the efficacy of application-specific storages, as CPU time is only measured for key creation and does not account for request queueing. It will show how quickly a key of the requested size can be created, regardless of whether it is stored in a single shared or application-dedicated queue. The S--Buffers have been modified to store keys in a queue data structure, where keys from the Q--Buffer are refformed into byte blocks, with each byte assigned a specific sequential identifier. When requested, keys are created simply by extracting a specific number of bytes from the queue. The sequential identifiers are exchanged between KMs during the supply key creation message~\cite{dervisevic2024large} and are used to ensure queue synchronization. Because keys are refformed in byte blocks in queues, the default key size (which represents the size of keys stored in the Q--Buffer) should not impact performance. 
Table~\ref{tab:storage-benchmark} displays the experiment-related configuration parameters. The experiment is repeated 1000 times. Because of the simplicity of the key creation procedure, it was expected that this key storage design would provide better overall performance than hash tables and larger key sizes. However, the results (see Section~\ref{sec:results}) have shown otherwise.

\subsection{Novel key storage design using deques}
\label{subsec:key-storage-benchmark}
To evaluate the effectiveness of our novel key storage design, we incorporate the concept of application-shared deques into QKDNetSim. When a new application requests keys from the QKD network services, the system checks existing deques to meet the request. If a deque of the requested size does not exist, and there is no existing deque with a smaller size that can fulfill the request, a new deque is created. Each deque is assigned a unique identifier in UUID format, allowing the key manager to communicate and track operations performed on a given deque easily. This enables efficient management and coordination of key storage operations within the QKD network. These deques are populated with key material from the encryption storages, facilitated by a modified fill procedure within QKDNetSim for the KM link.
Additionally, keys are assigned unique UUIDs before supply and synchronized over the KM link using a supply key creation message~\cite{dervisevic2024large}. The novel key storage design is expected to result in a more effective key supply than previously discussed key storage methods. This expectation is based on the optimized organization and management of key material within the application-shared deques. This allows for efficient allocation and retrieval of keys as needed by various cryptographic applications in the QKD network. Table~\ref{tab:storage-benchmark} displays the experiment-related configuration parameters. The experiment is repeated 1000 times. As with previous experiments, the CPU time for key creation is measured and used as a metric.

%TODO include a missing table, and include a laptop specifications used to run simulations

%\textbf{Ja bih ovdje ubacio zasebnu sekciju Simulation Setup i objasnio koja verzije QKDNetSim-a je korištena. Koja topologija je korištena, koliko čvorišta, koliko KMS-ova, da li je bilo rutiranje uključeno...}

\section{Results}
\label{sec:results}

\begin{figure} [h]
    \centering
    \begin{subfigure}{0.48\textwidth}
        \includegraphics[width=\textwidth]{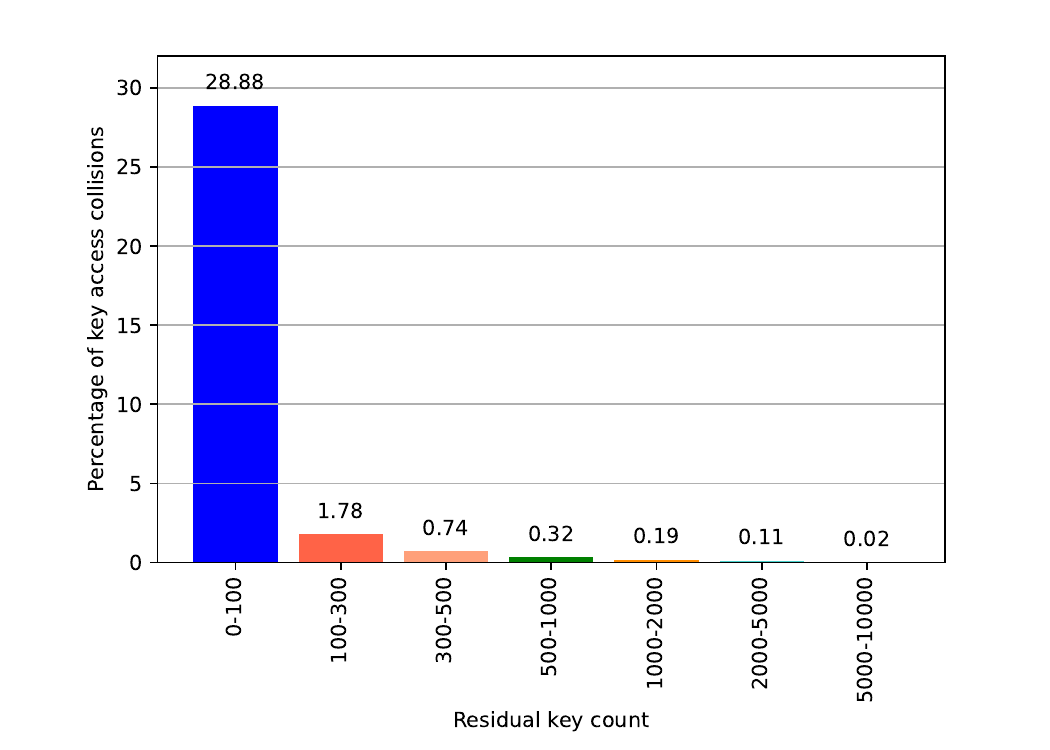}
        \caption{\label{fig:collision-count}}
    \end{subfigure}
    \hfill
    \begin{subfigure}{0.48\textwidth}
        \includegraphics[width=\textwidth]{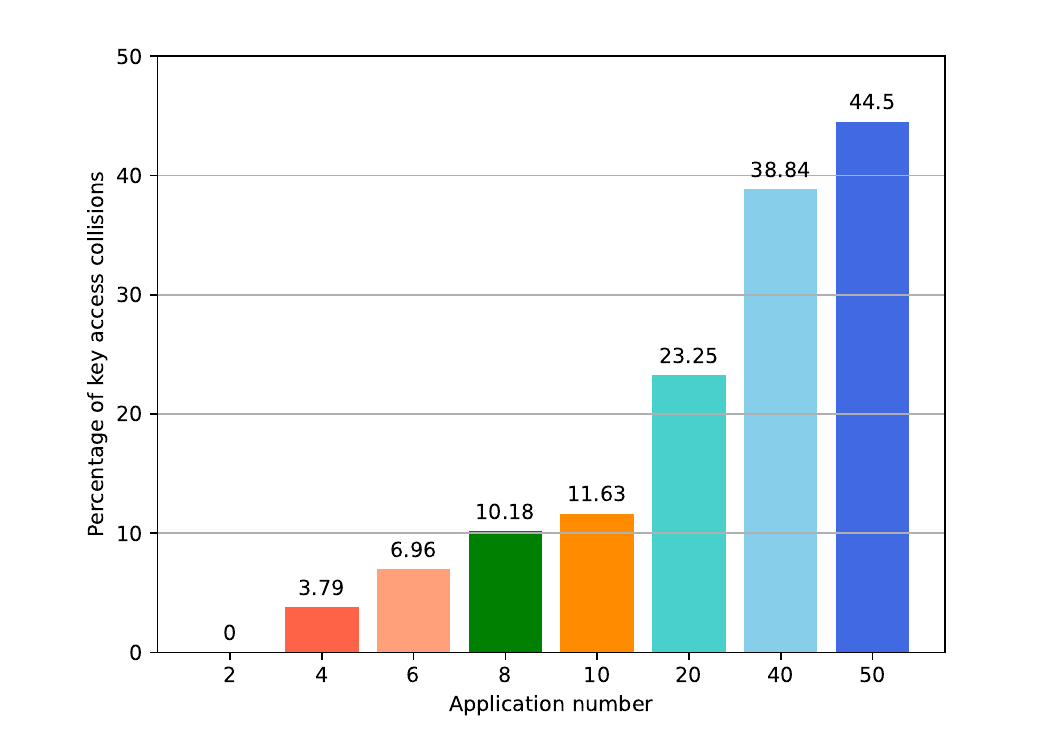}
        \caption{\label{fig:collision-app}}
    \end{subfigure}

    \caption{a) The percentage of key access collisions in different segments of the remaining number of keys.; b) The percentage of key access collisions for different number of concurrent applications running.}
    \label{fig:collisions-results}
\end{figure}

Figure~\ref{fig:collision-count} depicts the percentage of key access collisions within different intervals of residual key numbers available in a single common store. The figure supports the hypothesis that the residual key count strongly influences the frequency of key access collisions. When the residual key count is low, the frequency of key access collisions increases significantly. Furthermore, the high number of collisions causes applications to deprive the storage, making recovery more difficult. However, when the residual key count exceeds 1000 keys, the number of key access collisions becomes negligible. In the long run, it could still be a noticeable waste of limited key resources.

Figure~\ref{fig:collision-app} depicts the percentage of key access collisions for different numbers of concurrent applications running. Data is collected when the residual key count ranges from 10 to 100 keys. It displays a percentage relationship between the number of collisions and successful key accesses. The hypothesis presented in Section~\ref{sec:methodology} that the number of concurrent applications affects the frequency of key access collisions is correct. Increasing the number of applications causes more key access collisions. This observation is notable, particularly considering that the performance of QKD is still trailing behind the overall demands (i.e., there is a significant probability that the residual key count is low). Moreover, given the high cost associated with this technology, there is a notable shift towards enterprise-oriented applications (i.e., thousands of applications are connected to a single QKD node and share access to keys).

It should be noted that cryptographic applications in the given experiments obtain one key through a single GET\_KEY query. However, the ETSI method allows applications to request multiple keys at once. Consequently, this could impact the frequency of key access collisions, as the demands of the cryptographic application would be perceived as emanating from multiple applications.

\begin{figure} [h]
    \centering
    \begin{subfigure}{0.48\textwidth}
        \includegraphics[width=\textwidth]{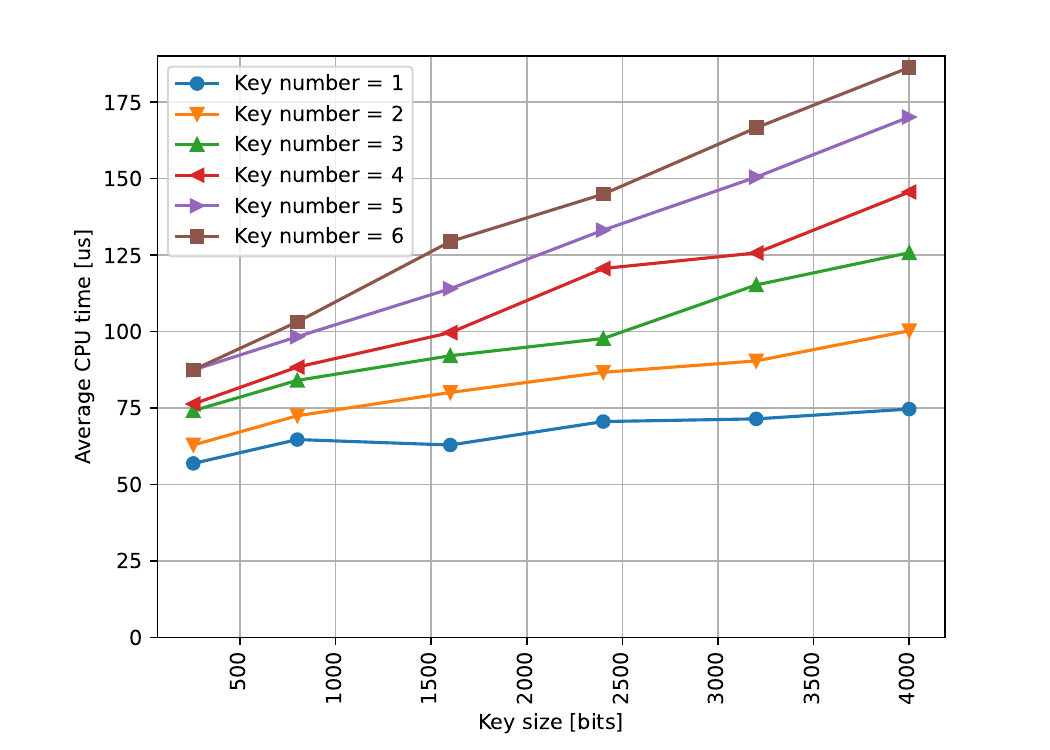}
        \caption{\label{fig:enc-dec-avg-cpu}}
    \end{subfigure}
    \hfill
    \begin{subfigure}{0.48\textwidth}
        \includegraphics[width=\textwidth]{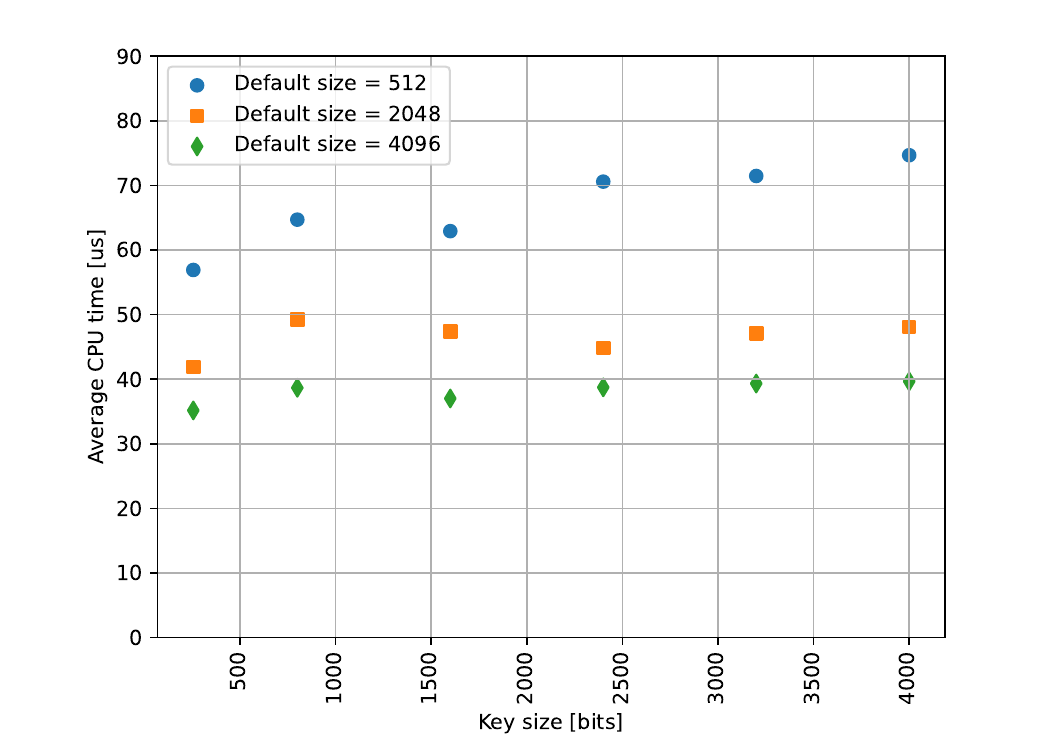}
        \caption{\label{fig:enc-dec-avg-cpu-variance}}
    \end{subfigure}

    \caption{a) The average CPU time in microseconds for creating a required number of keys of the requested size. The key storage design includes encryption and decryption storages with default key size of 512 bits; b) The average CPU time in microseconds for creating a key of the requested size in case of different default key sizes. The key storage design includes encryption and decryption storages.}
    \label{fig:original-results}
\end{figure}

Figure~\ref{fig:enc-dec-avg-cpu} depicts the average CPU time for creating a key of the requested size when the key storage design includes encryption and decryption storages. The required CPU time increases with the requested key size, as multiple keys are searched from storage to generate a required key. Increasing the number of keys requested via a single HTTP query results in increased CPU time. This can be viewed as a request for a larger single key, but it also includes additional processing, such as the generation and assignment of multiple UUIDs. Figure~\ref{fig:enc-dec-avg-cpu-variance} depicts how the default key size of the storage affects the average CPU time. The hypothesis that the default key size will significantly impact the average CPU time is correct. Increasing the default key size reduces average CPU time because fewer keys are searched and pulled from storage to generate a key of the required size.

\begin{figure} [h]
    \centering
    \begin{subfigure}{0.48\textwidth}
        \includegraphics[width=\textwidth]{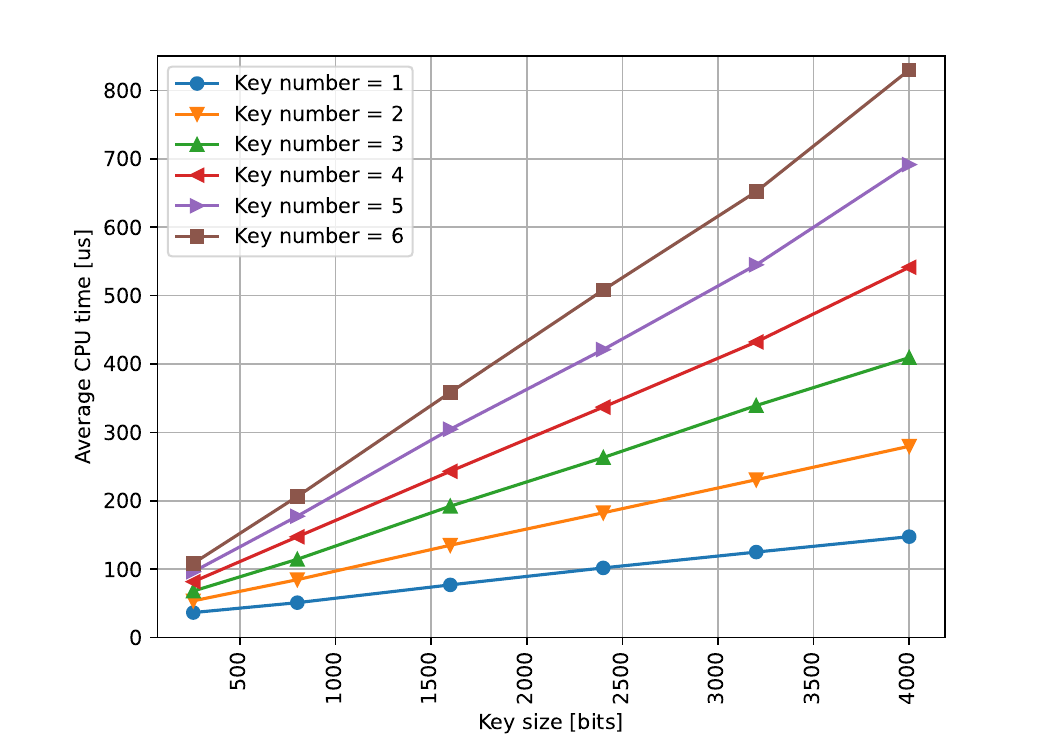}
        \caption{\label{fig:queue-avg-cpu}}
    \end{subfigure}
    \hfill
    \begin{subfigure}{0.48\textwidth}
        \includegraphics[width=\textwidth]{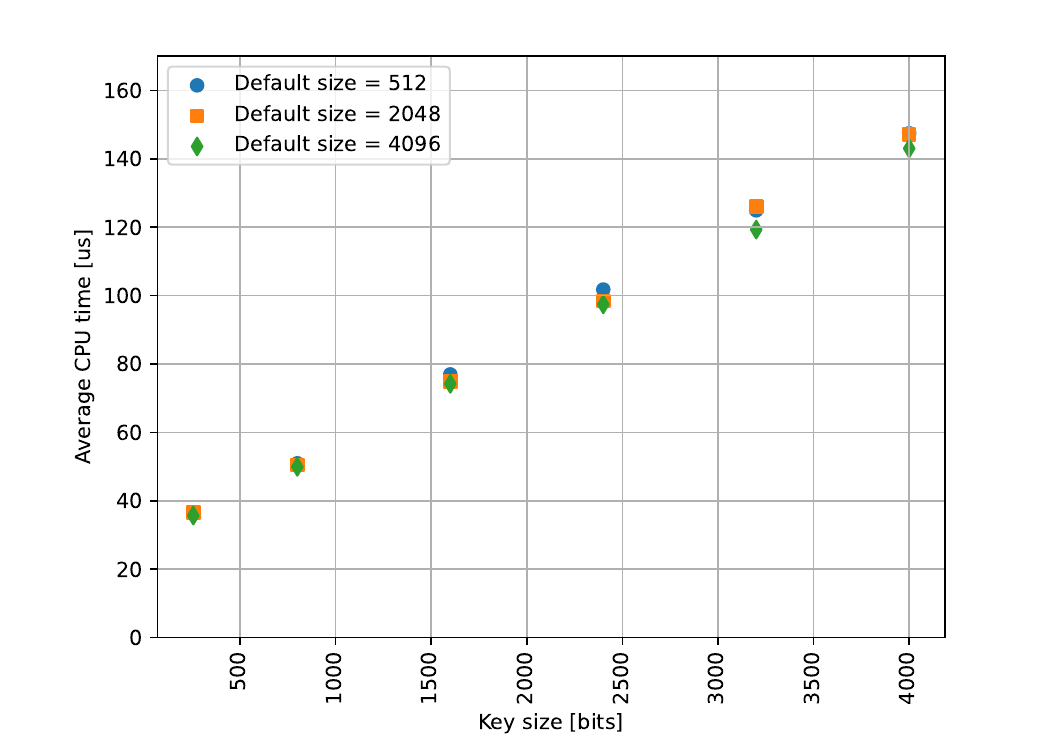}
        \caption{\label{fig:queue-avg-cpu-variance}}
    \end{subfigure}

    \caption{a) The average CPU time in microseconds for creating a required number of keys of the requested size. The key storage design includes purpose--based queues that store keys in byte blocks. The results are obtained for the default key size of 512 bits for the common storage; b) The average CPU time in microseconds for creating a key of the requested size in case of different default key sizes. The key storage design includes purpose--based queues that store keys in byte blocks.}
    \label{fig:queue-results}
\end{figure}

Figure~\ref{fig:queue-avg-cpu} depicts the average CPU time for creating a key of the requested size when the key storage design includes a purpose--based key storage that store keys in byte blocks. When a small number of keys of small size are requested, the achieved performance exceeds that of the previously analyzed key storage design. However, increasing key size, or key number, significantly increases average CPU time, and the design performs substantially worse than key storage designs with encryption and decryption key storage. This is because it takes many bytes to create a key of the required size. The design allows for far less complex key creation and instructions exchanged via the KM link. Furthermore, it is unaffected by the default key size (see Figure~\ref{fig:queue-avg-cpu-variance}) because keys are transformed into byte blocks when assigned to storage.

\begin{figure} [h]
    \centering
    \begin{subfigure}{0.48\textwidth}
        \includegraphics[width=\textwidth]{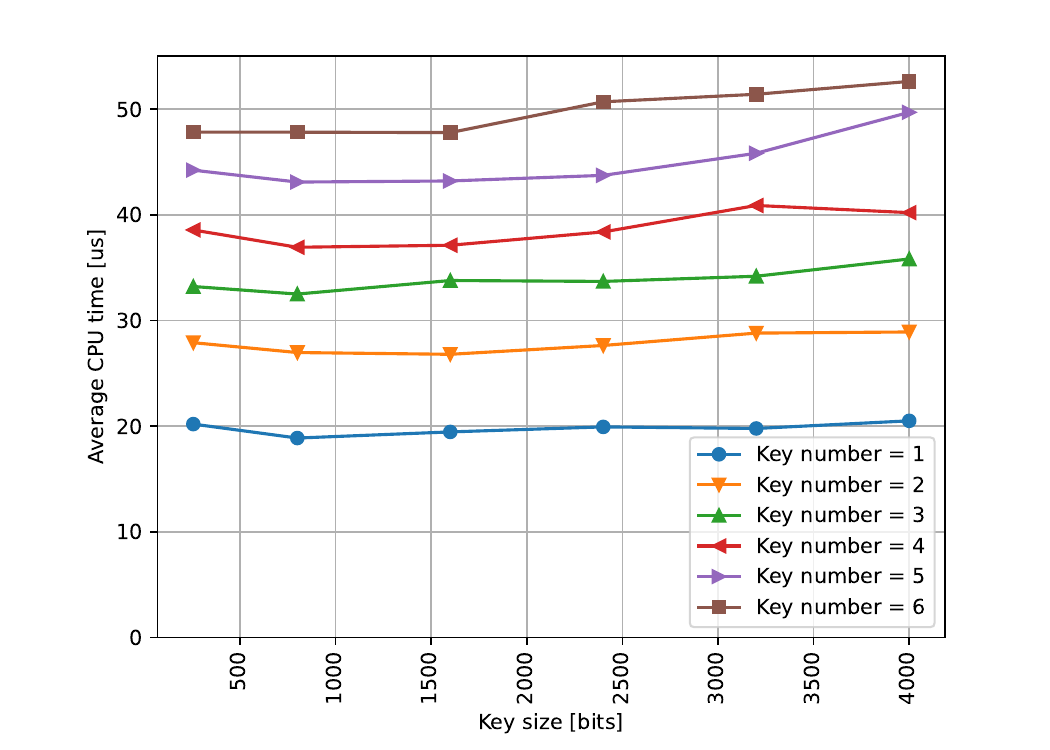}
        \caption{\label{fig:deque-avg-cpu}}
    \end{subfigure}
    \hfill
    \begin{subfigure}{0.48\textwidth}
        \includegraphics[width=\textwidth]{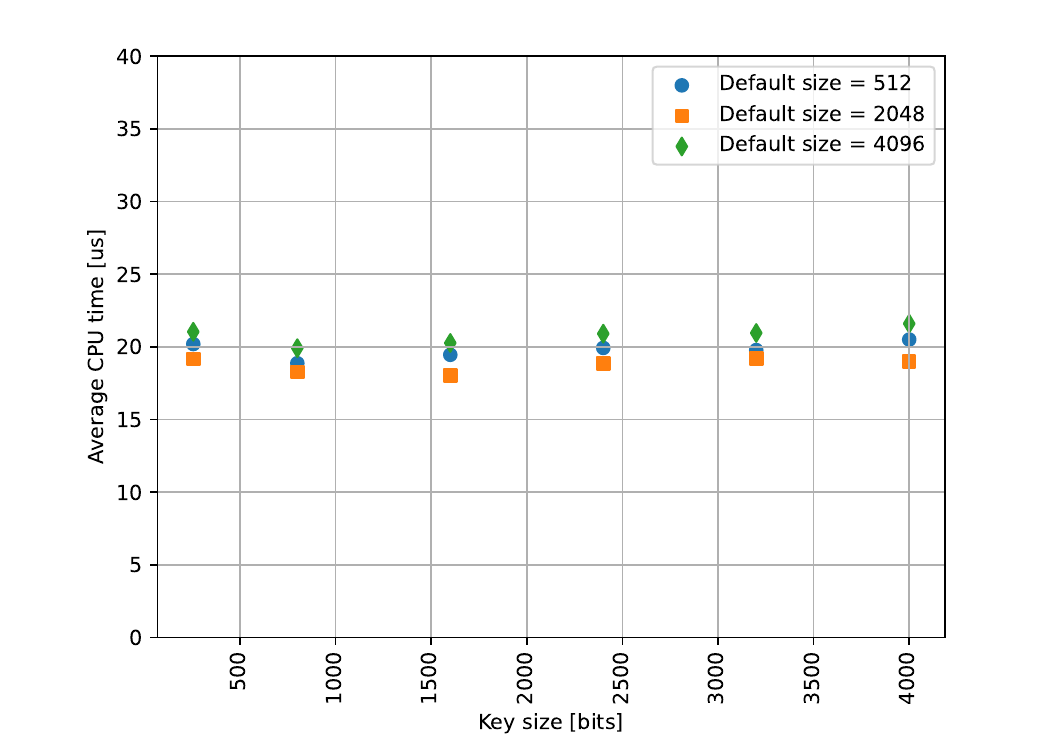}
        \caption{\label{fig:deque-cpu-variance}}
    \end{subfigure}

    \caption{a) The average CPU time in microseconds for creating a required number of keys of the requested size. The key storage design includes application--shared deques. The results are obtained for the default key size of 512 bits for the common storage; b) The average CPU time in microseconds for creating a key of the requested size in case of different default key sizes. The key storage design includes application--shared deques.}
    \label{fig:deque-results}
\end{figure}

Figure~\ref{fig:deque-avg-cpu} depicts the average CPU time for creating a key of the requested size within the novel key storage design that uses application-shared deques. The average CPU time for supply key creation remains constant for varying key sizes. However, this almost constant value increases with the number of requested keys. The novel key storage design outperforms previously analyzed designs, with a lower average CPU time for key creation. The performance is unaffected by the default key size, as shown in Figure~\ref{fig:deque-cpu-variance}.

%%%%%%%%%%%%%%%%%%%%%%%%%%%%%%%%%%%%%%%%%%
\section{Discussion}
\label{sec:discussion}
This article demonstrates that key supply effectiveness is determined by key storage design. It examines two approaches derived from practical testbeds: key storage design, which includes encryption and decryption storage, and application-specific key storage design. Although the latter design was incorporated into encryption and decryption storages, the results are applicable to the original application-specific key storage design. 
The results show that when keys are organized in sequential byte format, the design's effectiveness rapidly degrades as key sizes increase. In contrast, the default key size significantly impacts storing keys as multiple-byte blocks and performing necessary merging and splitting operations. Our novel key storage design improves by utilizing application--shared deques established by the key size requirements. The average CPU time for supply key creation remains constant across key sizes and increases with the number of requested keys. A single deque is associated with multiple applications with compatible key size requirements. 
The results show that for an average number of applications of 8.51, there are 4.45 average deques established. The key manager successfully identifies and assigns compatible key size requirements to a shared deque. However, in our experiment, the key size requirements were mainly compatible. If the requested key sizes vary uniformly, the proposed key storage design would result in a large number of established deques, which would be difficult to manage. As a result, a hybrid approach to key storage design is required, including encryption and decryption key storages realized with hash tables, as well as designs that use application-shared deques. 
The key manager should monitor application requests and decide whether to create a deque with an appropriate key size that can be shared by multiple applications to improve key supply efficiency. Given their common use in AES encryption, it's reasonable to expect specific key sizes, like 256 bits, to be frequently requested. Additionally, the suggested key storage design could enhance key supply efficiency because there's a direct correlation between packet size and key size in OTP encryption, and certain packet sizes are more common due to application characteristics. Further improvements are possible, such as using deques with larger key sizes to fulfill requests for multiple keys. In such cases, a single key from the deque is divided equally into multiple keys based on the required key size. Furthermore, once established, the deques can be filled directly with keys from the common storage, lowering communication overhead on the KM link.

%%%%%%%%%%%%%%%%%%%%%%%%%%%%%%%%%%%%%%%%
%	CONCLUSION %180
%%%%%%%%%%%%%%%%%%%%%%%%%%%%%%%%%%%%%%%%
   
\section{Conclusion}
\label{sec:conclusion}
The article introduces a novel key storage design aimed at improving key management efficiency in QKD networks. This design involves application-shared key storages, which store pre-formatted keys of commonly requested sizes by multiple applications. The number of application-shared storages is minimized by fulfilling compatible key size requirements using a single shared storage whenever possible. The time taken to create supply keys remains constant regardless of the key size requirement, and it proves to be efficient compared to previous approaches. The approach presented represents a significant advancement in integrating and applying QKD services in critical infrastructures, such as 5G networks, where the timely delivery of keys is crucial.

\section*{Acknowledgment}

The research leading to the published results was supported by the Ministry of the  Interior of the  Czech  Republic under grant ID VJ01010008  within the project  Network  Cybersecurity in Post-Quantum  Era, partly by the NATO SPS G5894 project "Quantum Cybersecurity in 5G Networks (QUANTUM5)" and  also by the H2020 project OPENQKD under grant agreement No. 857156. This work was also supported by the Ministry of Science, Higher Education and Youth of Canton Sarajevo, Bosnia and Herzegovina under Grant No. 27-02-35-37082-42/23.

\bibliographystyle{unsrtnat}

\end{document}